\documentclass[aps,twocolumn,showpacs]{revtex4}

\usepackage{amsmath}
\usepackage{amsfonts}
\usepackage{amssymb}
\usepackage{color}

\def\bc{\begin{center}}

\def\ec{\end{center}}
\def\be{\begin{eqnarray}}
\def\ee{\end{eqnarray}}

\newcommand{\omits}[1]{}

\definecolor{dyellow}{rgb}{1.,0.8,.0}
\definecolor{myblue}{rgb}{.1,.1,.7}
\definecolor{dcyan}{rgb}{.0,.6,.6}
\definecolor{dmagenta}{rgb}{0.6,0.0,0.6}
\definecolor{brown}{rgb}{0.6,0.2,0.}
\definecolor{darkblue}{rgb}{.0,.0,0.5}
\definecolor{darkred}{rgb}{0.75,0.0,0.0}
\definecolor{orange}{rgb}{1.,.6,.0}
\definecolor{dorange}{rgb}{0.8,.4,.0}
\definecolor{darkgreen}{rgb}{0.0,0.6,0.0}
\definecolor{purple}{rgb}{.4,.0,.4}

\def\Si{\Sigma}

\def\dl{\delta}

\def\si{\sigma}

\def\d#1#2{\frac{\displaystyle #1}{\displaystyle #2}}

\newcommand\btd{\raise 2pt
\hbox{$\hat\bigtriangledown$}\hskip 1.5pt}
\newcommand\bt{\raise 2pt
\hbox{$\bigtriangledown$}\hskip 1.5pt}

\def\PRD{{\it Phys. Rev.}~{\bf D}}

\begin{document}



\title{On Beltrami Model of de Sitter Spacetime}


\author{{Han-Ying Guo}$^{1,2}$}
\email{hyguo@itp.ac.cn}
\author{{Chao-Guang Huang}$^{3}$}
\email{huangcg@mail.ihep.ac.cn}
\author{{Zhan Xu}$^{4}$}
\email{zx-dmp@mail.tsinghua.edu.cn}
\author{{Bin Zhou}$^{3}$} \email{zhoub@mail.ihep.ac.cn}
\affiliation{%
${}^1$ CCAST (World Laboratory), P.O. Box 8730, Beijing
   100080, China,}

\affiliation{%
${}^2$ Institute of Theoretical Physics,
 Chinese Academy of Sciences,
 P.O.Box 2735, Beijing 100080, China,}

\affiliation{%
${}^3$ Institute of High Energy Physics, Chinese Academy of
Sciences, P.O. Box 918-4, Beijing
   100039, China,}
\affiliation{%
${}^4$ Physics Department, Tsinghua University, Beijing
   100084, China,}
\date{November 7, 2003}


\begin{abstract}
Based on some  important properties of $dS$ space, we present a
Beltrami model ${\cal B}_\Lambda$ that may shed light on the
observable puzzle of $dS$ space and the paradox between the
special relativity principle and cosmological principle. In ${\cal
B}_\Lambda$, there are inertial-type coordinates and inertial-type
observers. Thus, the classical observables can be defined for test
particles and light signals. In addition, by choosing the
definition of simultaneity the Beltrami metric is transformed to
the Robertson-Walker-like metric. It is of positive spatial
curvature of order $\Lambda$. This is more or less indicated
already by the CMB power spectrum from WMAP and should be further
confirmed by its data in large scale.
\end{abstract}


\maketitle


\section{Introduction}
Among many puzzles in ordinary approach to $dS$ space \cite{SSVB},
one is how to define observables on it \cite{Str}. There  has been
also a long-standing paradox between the special relativity (SR)
principle and the cosmological principle \cite{BBR}, called the
SRP-CP paradox, since cosmological models were proposed and
especially since the cosmic microwave background (CMB) was
discovered.  Roughly speaking, it may be simply described as:
``what would happen for SR, if the CMB could be discovered before
1905?"

In this letter, we present a proposal to the $dS$ observable
puzzle and the SRP-CP paradox. The key observation is based upon
some simple but important properties of $dS$ space
\cite{Gursey}--\cite{Guo1}. In fact, among various  metrics of
$dS$ spaces, there is an important one in which $dS$ space is in
analog with Minkowski space. It is the $dS$ space with
Beltrami-like metric, called $BdS$ space and denoted as ${\cal
B}_\Lambda$.  $BdS$ space is precisely the Beltrami-like model
\cite{beltrami} of a 4-hyperboloid ${\cal S}_\Lambda$ in 5-d
Minkowski space, i.e. ${\cal B}_\Lambda \backsimeq {\cal
S}_\Lambda $. In  ${\cal B}_\Lambda$ there exist  a set of
Beltrami coordinate  systems, which covers ${\cal B}_\Lambda$
patch by patch, and in which test particles and light signals move
along the timelike and null geodesics, respectively, with {\it
constant} coordinate velocity. Therefore, they look like in free
motion in a spacetime without gravity.
Thus, the Beltrami coordinates and observers ${\cal O}_{B}$ at
these systems may be regarded as of ``global inertial-type". And
the classical observables for these particles and signals may be
well defined. Thus, it may shed light on the $dS$ observable
puzzle.

Why in a constant curvature spacetime do there exist such
motions and observers of global inertial-type?  %
As is well known, if we start with the 4-d Euclidean geometry and
weaken the fifth axiom, then there should be 4-d Riemann, Euclid,
and Lobachevski geometries at equal footing. Importantly, their
geodesics are globally straight lines in certain coordinate
systems that are just ones in analog with the coordinate systems
in Beltrami model of Lobachevski plane \cite{beltrami}, and under
corresponding transformation groups the systems transform among
themselves. Now changing the signature to $-2$, these constant
curvature spaces turn to $dS$, Minkowski, and $AdS$ spacetimes,
respectively, and those straight lines are classified by
timelike, null and spacelike straight world-lines. Thus, in analog
with SR, the former two should describe the ``global inertial-type
motions" for free particles and light signals, respectively. And,
Einstein's SR principle should also be available in ${\cal
B}_\Lambda$, called SR-type principle with respect to $dS$,
Poincar$\grave{e}$, $AdS$ group, respectively.

If the Beltrami coordinates make sense, they should concern with
the measurements in laboratory at one patch and
simultaneity should be defined with respect to the time
coordinate. On the other hand, the simultaneity may also be
defined by the proper time of a clock rest at the origin of
Beltrami spatial coordinates. Thus, there are two kinds of
simultaneity in ${\cal B}_\Lambda$. Consequently, if the
simultaneity is transformed from the first to the second, the
Beltrami metric is reduced to the Robertson-Walker(RW)-like metric
in ${\cal B}_\Lambda$ with positive, rather than zero or negative,
spatial curvature and its deviation from zero is in the order of
cosmological constant $\Lambda$. Thus, the second simultaneity
should link with the observation in the cosmological scale. This
important property differs with either SR where two of them
coincide, or usual cosmological models where $k$ is a free
parameter. Thus, this property sheds light on the SRP-CP paradox.
In addition, the tiny spatial closeness seems more or less to have
been already confirmed by the CMB power spectrum data from WMAP
\cite{WMAP} and should be further checked by its data in large
scale.

This letter is organized as follows.  In section 2, we briefly
set up a framework for ${\cal B}_\Lambda$.  In section 3 we study
the motion of particles and light signals and define their
observables. In section 4, we define two kinds of simultaneity,
the length of a ruler, and so on.  We also show that ${\cal
B}_\Lambda$ can be reduced to RW-like metric with a slightly
closed space. Finally, we end with a few remarks.

\section{The Beltrami-$dS$ Spacetime}

We start with a 4-d hyperboloid ${\cal S}_\Lambda$ embedded in a
5-d Minkowski space with $\eta_{AB}= {\rm diag}(1, -1, -1, -1,
-1)$:
 \be\label{5sphr}%
 {\cal S}_\Lambda:  &&\eta^{}_{AB} \xi^A \xi^B= -R^2,%
\\ %
\label{ds2}%
&&ds^2=\eta^{}_{AB} d\xi^A d\xi^B , 
\ee where $R^2:=3\Lambda^{-1}$ and $A, B=0, \ldots, 4$.  Clearly,
Eqs. (\ref{5sphr}) and (\ref{ds2}) are invariant under $dS$ group
${\cal G}_\Lambda =SO(1,4)$.

 The Beltrami coordinates
are defined patch by patch on ${\cal B}_\Lambda\simeq{\cal
S}_\Lambda$. For intrinsic geometry of ${\cal B}_\Lambda$, there
are at least eight patches $U_{\pm\alpha}:= \{ \xi\in{\cal
S}_\Lambda :  \xi^\alpha\gtrless 0\}, \alpha=1,
\cdots, 4$. In $U_{\pm 4}$, for instance, the Beltrami
coordinates are
\be \label{u4}%
&&x^i|_{U_{\pm 4}} =R {\xi^i}/{\xi^4},\qquad i=0,\cdots, 3;  \\
&&\xi^4=\pm({\xi ^0}^2-\sum _{a=1}^{3}{\xi ^a}^2+ R^2 )^{1/2} \neq
0.
\ee%
In the patches $\{U_{\pm a}, a=1,2,3\}$,
\begin{equation}  %
y^{j'}|_{U_{\pm a}}=R{\xi^{j'}} /{\xi^a},\quad
j'=0,\cdots,\hat{a}\cdots,4; \quad \xi^{a}\neq 0,
\end{equation}
where $\hat{a}$ means omission of $a$. It is important that all
transition functions in  intersections are of ${\cal G}_\Lambda$,
say, in $U_4\bigcap U_3$, the transition function $T_{4,3}
=\xi^3/\xi^4=x^3/R=R/y^4 \in {\cal G}_\Lambda$ so that
$x^i=T_{4,3}y^{i'} $.

In each patch, there are condition and Beltrami metric
\begin{eqnarray}\label{domain}
\sigma(x)&=&\sigma(x,x):=1-R^{-2}
\eta_{ij}x^i x^j>0,\\
\label{bhl} ds^2&=&[\eta_{ij}\sigma(x)^{-1}+ R^{-2}
\eta_{ik}\eta_{jl}x^k x^l \sigma(x)^{-2}]dx^i dx^j.
\end{eqnarray}
 Under fractional linear
transformations of ${\cal G}_\Lambda$
\begin{equation}\label{G}
\begin{array}{l}
x^i\rightarrow \tilde{x}^i=\pm\sigma(a)^{1/2}\sigma(a,x)^{-1}(x^j-a^j)D_j^i,\\
D_j^i=L_j^i+{ R^{-2}}%
\eta_{jk}a^k a^l (\sigma(a)+\sigma(a)^{1/2})^{-1}L_l^i,\\
L:=(L_j^i)_{i,j=0,\cdots,3}\in SO(1,3),
\end{array}\end{equation}
where $\eta_{ij}%
={\rm diag} (1, -1,-1,-1)$ in $U_{\pm\alpha}$, Eqs. (\ref{domain})
and (\ref{bhl}) are invariant.

 Note that Eqs. (\ref{domain})-(\ref{G}) are defined on ${\cal
B}_\Lambda$ patch by patch.  This is, in fact, a cornerstone for
the SR-type principle. $\sigma(x)=0$ may be considered as the
boundary of $BdS$ spacetime, $\partial{\cal B}_\Lambda$.

For two separate events $A(a^i)$ and $X(x^i)$ in ${\cal
B}_\Lambda$,
\be\label{lcone0} %
{\Delta}_\Lambda^2(A, X) = R
[\sigma^{-1}(a)\sigma^{-1}(x)\sigma^2(a,x)-1]
\ee %
is invariant under ${\cal G}_\Lambda$. Thus, the interval
between $A$ and $B$ is timelike, null, or spacelike,
respectively, according to%
\begin{equation}\label{lcone}%
\Delta_\Lambda^2(A, B)\gtreqqless 0.%
\end{equation}

The proper length of timelike or spacelike between $A$ and $B$ are
integral of ${\cal I} ds$ over the geodesic segment
$\overline{AB}$:
\be \label{AB1}%
S_{timelike}(A,B)&=&R \sinh^{-1} (|\Delta(a,b)|/R), \\
\label{AB1sl} S_{spacelike}(A,B)&=& R \arcsin (|\Delta(a,b)|/R),
\ee
where ${\cal I}=1, -i$ for timelike or spacelike,
respectively.

It can be shown that the light-cone at $A$ with running points
$X$ is
\be \label{nullcone} %
{\cal F}_{\Lambda}:= R
\{\sigma(a,x) \mp [\sigma(a)\sigma(x)]^{1/2}\}=0.%
 \ee%
It satisfies the null-hypersurface condition
 \be\label{Heqn}%
\left . g^{ij}\frac{\partial f}{\partial x^i}\frac{\partial
f}{\partial x^j}\right |_{f=0}=0, %
\ee
where $g^{ij}=\sigma(x)(\eta^{ij}-R^{-2} x^i x^j)$ is the inverse
Beltrami metric.

\section{Motion and Observables of Test Particles}
We now show that in ${\cal B}_\Lambda$ the geodesics are
Lobachevski-like straight world lines, along which the observables
for test particles and signals may be well defined.

For a free particle with mass $m_{\Lambda 0}$, it should move
globally along a timelike geodesic with respect to the Beltrami
metric. It is easy to show that the geodesic equation is
equivalent to
\begin{equation}\label{pi}
\frac{dp^i}{ds}=0, \quad
p^i:=m_{\Lambda 0}\sigma(x)^{-1}\frac{dx^i}{ds}=C^i={\rm const.}
\end{equation}
This implies that under the initial condition
\[
x^i(s=0)=b^i, \qquad \frac{dx^i}{ds}(s=0)=c^i
\]
with the constraint
\[
g_{ij}(b)c^i c^j=1,
\]
a new parameter $w = w(s)$ can be chosen such that
the geodesic is just a straight world-line
\begin{equation}
x^i(w)=c^iw+b^i.
\end{equation}
This property is in analog with the straight line in the Beltrami
model of Lobachevski plane.
The parameter $w$ can be integrated out,
\be \label{w1}
w(s) = \begin{cases}
R e^{\mp s/R}\sinh \d s R,   &    \eta_{ij}\,c^i c^j = 0, \medskip \cr
\d {R\sinh \frac s R} {\frac {\eta_{ij}\,c^i b^j}{R\sigma(b)}\sinh \frac s R
    + \cosh \frac s R},  & \eta_{ij}\,c^i c^j \neq  0.
\end{cases}
\ee

Similarly, a light signal moves globally along a null geodesic.
The null geodesic equation formally still has the first
integration
\begin{equation}
\sigma ^{-1}(x)\frac{dx^i}{d\tau }={\rm constant},
\end{equation}
 but now the condition $ds = 0$ also holds, where
$\tau $ is an affine parameter.  Again, under the initial
condition
\be %
x^i(\tau =0)=b^i,  \qquad \d {dx^i}{d\tau }(\tau =0)=c^i. %
\ee
and the constraint \be g_{ij}(b)\,c^ic^j=0, \ee the null
geodesic can be expressed as a straight line
\[
x^i = c^i w(\tau) +b^i,
\]
where
\be \label{w2}%
w(\tau) = \begin{cases} {~ \tau}, & \eta _{ij}\,c^ic^j =0, \cr -
\d { R^2\sigma (b)}{|\eta _{ij}c^ic^j|}\left( \d 1{\tau +\tau
_0}-\d 1{\tau _0}\right), & \eta _{ij}\,c^ic^j \neq 0,
\end{cases}
\ee
with
\begin{equation}\nonumber
\tau _0=\sqrt{\frac{ R^2 \sigma (b)}{|\eta _{ij}c^ic^j|}}.
\end{equation}

Thus, for free particle and light signal the components of the
coordinate velocity are constants, respectively:
\begin{equation}\label{vi}
\frac{dx^a}{dt}=v^a;\quad \frac{d^2x^a}{dt^2}=0;\quad a=1,2,3.
\end{equation}
Of course, this makes sense only if the Beltrami coordinate system
is of physical meaning as inertial-type.

Now we are ready to define the observables for test particles.
From the (\ref{pi}), it is natural to define the conservative
quantities $p^i$ along the geodesic as the 4-momentum of a free
particle with mass $m_{\Lambda, 0}$ and its zeroth component as
the energy. Note that this 4-momentum is no longer a 4-vector
rather a pseudo 4-vector.

Furthermore, for a free particle a set of quantities $L^{ij}$ may
also be defined by the following equation and they are also
conserved along a geodesic
\begin{equation}\label{angular4}
L^{ij}=x^ip^j-x^jp^i;\quad \frac{dL^{ij}}{ds}=0.
\end{equation}
These may be called the 4-angular-momentum of a particle and they
are also no longer the components of an anti-symmetric tensor but
a pseudo anti-symmetric tensor.  However, $p^i$ and $L^{ij}$
constitute a 5-d angular momentum for a free
 particle in ${\cal S}_\Lambda$ and it is conserved along
the geodesics that are the straight world-line
\begin{equation}\label{angular5}
{\cal L}^{AB}:=m_{\Lambda
0}(\xi^A\frac{d\xi^B}{ds}-\xi^B\frac{d\xi^A}{ds}); \quad
\frac{d{\cal L}^{AB}}{ds}=0.
\end{equation}
And Einstein's famous formula for such a kind of free particles
can be generalized in ${\cal B}_\Lambda$:
\begin{equation}\label{eml}
\frac{\lambda}{2}{\cal L}^{AB}{\cal L}_{AB}=E^2-{\bf P}\,^2-
\frac{1}{R^2}{\bf L}^2=m_{\Lambda 0}^2,
\end{equation}
where ${\cal L}_{AB}=\eta_{AC}\eta_{BD}{\cal L}^{CD}$, $m_{\Lambda
0}$ introduced above should be the inertial-type mass for a free
 particle. It is well defined together with the energy, momentum and angular momentum at classical level.

It can further be shown that $m_{\Lambda 0}$ is the eigenvalue of
the first Casimir operator of the $dS$ group ${\cal G}_\Lambda$
\cite{Gursey}.
In addition, spin 
can also be well defined as it was done in the relativistic
quantum mechanics in Minkowski spacetime. We will explain the
issue in detail elsewhere.

In any case, these offer a consistent way to define the
observables for free particles and this kind of definitions differ
from any others in $dS$ space. Of course, these issues
significantly indicate that the motion of a free particle in
${\cal B}_\Lambda$ should be of inertial-type in analog to
Newton's and Einstein's conception for the inertial motion of a
free particle with constant velocity. Consequently, the coordinate
systems with Beltrami metric should be the globally inertial-type
systems and corresponding observer at the origin of the system
should be of inertial-type as well.

\section{On Definitions of Simultaneity, Space-Time Measurements and
Robertson-Walker-like Metric} Among physical measurements,
space-time measurements are most fundamental.  In order to make
space-time measurements, one should first define simultaneity.  It
is worth to note that there are two different kinds of definitions
of simultaneity in the ${\cal B}_\Lambda$.  Now, we discuss the two kinds of
definitions separately.

In SR, coordinates have measurement significance that is linked
with the SR principle. Namely, the difference in time coordinate
stands for the time interval, and the difference in spatial
coordinate stands for the spatial distance. Similar to Einstein's
SR, one can define that two events $A$ and $B$ are simultaneous if
and only if the Beltrami time coordinate $x^0$ for the two events
are same, 
\be %
a^0:=x^0(A) =x^0(B)=:b^0. %
\ee
It is with respect to this simultaneity that free particles move
along straight lines with uniform velocities.  The simultaneity
defines a 3+1 decomposition of spacetime
\be
ds^2 =  N^2 (dx^0)^2 - h_{ab} \left (dx^a+N^a dx^0 \right )
\left (dx^b+N^b dx^0 \right ) %
\ee
with the lapse function, shift vector, and induced 3-geometry on
3-hypersurface $\Si_c$ in one coordinate patch.
\begin{eqnarray}
& & N=\{\si_{\Si_c}(x)[1-(x^0 /R)^2]\}^{-1/2}, \nonumber \\%
& & N^a=x^0 x^a[ R^2-(x^0)^2]^{-1},
\nonumber \\
& & h_{ab}=\dl_{ab} \si_{\Si_c}^{-1}(x)-{ [R\si_{\Si_c}(x)]^{-2}
\dl_{ac} \dl_{bd}}x^c x^d ,
\end{eqnarray}
respectively, where $\si_{\Si_c}(x)=1-(x^0{/R})^2 + {\dl_{ab}x^a
x^b /R^2}$,  $\dl_{ab}$ is the Kronecker $\dl$-symbol,
$a,b=1,2,3$.  In particular, at $x^0=0$, $\si_{\Si_c}(x)=1+{
\dl_{ab} x^a x^b/R^2}$. In a vicinity of the origin of Beltrami
coordinate system in one patch, 3-hypersurface $\Si_c$ acts as a
Cauchy surface.

This simultaneity defines the laboratory time in one
patch. According to the spirit of SR as well as the SR-type
principle, the Beltrami coordinates define, in such a
manner, the standard clocks and standard rulers in laboratory on
${\cal B}_\Lambda$. To measure the time of a process or the size
of an object, we just need to compare with Beltrami coordinates.

There is another simultaneity, however. It is, in fact, with
respect to the proper time of a clock rest at spatial origin of
the Beltrami coordinate system. It can be shown that the proper
time $\tau_{\Lambda>0}$ of a rest clock on the time axis of
Beltrami coordinate system, $\{x^a=0\}$, reads
\begin{eqnarray}\label{ptime}
\tau_{\Lambda>0}=R \sinh^{-1} (R^{-1}\sigma^{-\frac{1}{2}}(x)x^0).
\end{eqnarray}%
Therefore, we can define that the events are simultaneous with
respect to the proper time of a clock rest at the origin of the
Beltrami spatial coordinates if and only if
\begin{equation}\label{smlt}
x^0\sigma^{-1/2}(x,x)=\xi^0:=R \sinh(R^{-1}\tau)=\rm constant.
\end{equation}
 The line-element %
on the simultaneous 3-d
hypersurface, denoted by ${\Sigma_\tau}$, can be defined as
\begin{equation}\label{dl}
dl^2=-ds^2_{\Sigma_\tau},%
\end{equation}
where
\be \begin{array}{l}\label{spacelike}
ds^2_{\Sigma_\tau} = R_{\Si_\tau}^2%
dl_{{\Sigma_\tau} 0}^2, \\
R_{\Sigma_\tau}^2%
:=\sigma^{-1}(x,x)\sigma_{\Sigma_\tau}(x,x)%
=1+ (\xi^0/R)^2,\\%
\sigma_{\Sigma_\tau}(x,x):=1+R^{-2}\delta_{ab}x^a x^b
>0, \\ 
dl_{{\Sigma_\tau} 0}^2:={
\{\delta_{ab}\sigma_{\Sigma_\tau}^{-1}(x)
-[R\sigma_{\Sigma_\tau}(x)]^{-2}\delta_{ac}\delta_{bd}x^c x^d\}}
 dx^a dx^b.
\end{array}\ee
Note that ${\Sigma_\tau}$ has a positive spatial curvature.

It should be pointed out that this simultaneity is closely linked
with the cosmological principle. In fact,
 it is significant that if $\tau_{\Lambda>0}$ is taken as
a ``cosmic time", the Beltrami metric (\ref{bhl}) becomes
\begin{equation}\label{dsRW}
ds^2=d\tau^2-dl^2=d\tau^2-R^2\cosh^2( R^{-1}\tau) dl_{{\Sigma_\tau} 0}^2.%
\end{equation}
It is the RW-like metric with a positive spatial curvature and the
simultaneity is globally defined in whole ${\cal B}_\Lambda$.

It should be emphasized that the two definitions of simultaneity
do make sense in different kinds of measurements. The first
concerns the measurements in a laboratory and is related to the
SR-type principle in ${\cal B}_\Lambda$ patch by patch,
while the second concerns the observations with cosmological
principle. Furthermore, the relation between the Beltrami metric
and its RW-like correspondence (\ref{dsRW}) is meaningful. It
links the coordinate time $x^0$ in laboratory and the cosmic time
$\tau$ in a manifest way. This may shed light on the puzzle
between laboratory time coordinate and cosmic time with arrow,
another version of the SRP-CP puzzle. This also shows
that the 3-d cosmic space is $S^3$ rather than flat.  The
deviation from the flatness is of order
$\Lambda$. Obviously, this spatial closeness of the universe is
another remarkable property different from the standard
cosmological model with flatness. This property seems more or less
already indicated by the CMB power spectrum from WMAP \cite{WMAP}
and should be further checked by its data in large scale.

\section{Remarks}
We have shown that in ${\cal B}_\Lambda$, the Beltrami coordinates
should be regarded as inertial-type since the test particles and
signals move along the timelike, null straight world lines,
respectively. Therefore, their classical observables can be well
defined. Consequently, the $dS$ observable puzzle %
might be solved at least partly in such a way.

Furthermore, it is meaningful that the SRP-CP paradox might be
solved by the relation between the Beltrami metric and the RW-like
metric in ${\cal B}_\Lambda$. This relation links the coordinate
time $x^0$ in the laboratory of one patch and the cosmic time
$\tau$ in the large  scale and shows that the 3-d cosmic space is
slightly closed.  This should be further checked by the data from
WMAP \cite{WMAP} in large scale.

In fact all properties in ${\cal B}_\Lambda$ are in
analog with SR and coincide with it if $\Lambda=0$.  Meanwhile,
the SRP-CP paradox and other puzzles will appear again.



\begin{acknowledgments}
The authors would like to thank Professors Z. Chang, Q. K. Lu, J.
Z. Pan, S. K. Wang, K. Wu and C. J. Zhu for valuable discussion.
This work is partly supported by NSFC { under Grants Nos.
90103004, 10175070, 10375087.}
\end{acknowledgments}

\end{document}